\documentclass[conference]{IEEEtran}
\IEEEoverridecommandlockouts
\usepackage{cite}
\usepackage{amsmath,amssymb,amsfonts}
\usepackage{graphicx, color}
\usepackage{subcaption}
\usepackage[export]{adjustbox}
\usepackage{textcomp}
\usepackage{url} 
\usepackage[table]{xcolor} 
\usepackage{paralist}
\usepackage{enumitem}
\usepackage{multirow}
\usepackage{hyperref}
\usepackage{algorithmic}
\usepackage[linesnumbered,ruled,vlined]{algorithm2e}
\hypersetup{
    colorlinks=true,
    linkcolor=red, 
    citecolor=blue, 
    urlcolor=magenta    
}
\def\ie{{\em i.e.}}
\def\eg{{\em e.g.}}
\def\BibTeX{{\rm B\kern-.05em{\sc i\kern-.025em b}\kern-.08em
    T\kern-.1667em\lower.7ex\hbox{E}\kern-.125emX}}

\newcommand{\smallurl}[1]{\footnotesize\url{#1}}

\definecolor{baselinecolor}{gray}{.9}
\newcommand{\baseline}[1]{\cellcolor{baselinecolor}{#1}}

\def\sign{\texttt{sign}}

\begin{document}

\title{Robust COVID-19 Detection in CT Images with CLIP}

\author{\IEEEauthorblockN{ Li Lin\textsuperscript{1}$^+$\thanks{$^+$Co-first Authors}, Yamini Sri Krubha\textsuperscript{1}$^+$,  Zhenhuan Yang\textsuperscript{2}, Cheng Ren\textsuperscript{3}, Thuc Duy Le\textsuperscript{4}, Irene Amerini\textsuperscript{5}, Xin Wang\textsuperscript{3},\\ 
Shu Hu\textsuperscript{1}$^*$\thanks{$^*$Corresponding Author} }
\IEEEauthorblockA{
{\textsuperscript{1}Purdue University, {\tt \small \{lin1785, ykrubha,  hu968\}@purdue.edu} }\\
\textsuperscript{2}Etsy, Inc, Brooklyn, New York, USA {\tt \small zhenhuan.yang@hotmail.com}\\
\textsuperscript{3}University at Albany, State University of New York {\tt \small \{cren, xwang56\}@albany.edu}\\
\textsuperscript{4}University of South Australia {\tt \small thuc.le@unisa.edu.au}\\
\textsuperscript{5}Sapienza University of Rome {\tt \small amerini@diag.uniroma1.it}}
}

\maketitle
\thispagestyle{plain}
\pagestyle{plain}

\begin{abstract}
In the realm of medical imaging, particularly for COVID-19 detection, deep learning models face substantial challenges such as the necessity for extensive computational resources, the paucity of well-annotated datasets, and a significant amount of unlabeled data. In this work,  we introduce the first lightweight detector designed to overcome these obstacles, leveraging a frozen CLIP image encoder and a trainable multilayer perception (MLP). Enhanced with Conditional Value at Risk (CVaR) for robustness and a loss landscape flattening strategy for improved generalization, our model is tailored for high efficacy in COVID-19 detection. Furthermore, we integrate a teacher-student framework to capitalize on the vast amounts of unlabeled data, enabling our model to achieve superior performance despite the inherent data limitations. Experimental results on the COV19-CT-DB dataset demonstrate the effectiveness of our approach, surpassing baseline by up to \textbf{10.6\%} in `macro' F1 score in supervised learning. The code is available at \smallurl{https://github.com/Purdue-M2/COVID-19_Detection_M2_PURDUE}.

\end{abstract}


\section{Introduction}
COVID-19 detection \cite{alazab2020covid} based on 3-D chest CT scans is a diagnostic technique that uses computed tomography (CT) imaging to capture detailed images of the lungs and chest area in three dimensions. This method has been explored and utilized during the COVID-19 pandemic as a means to detect and assess the severity of infections caused by the SARS-CoV-2 virus. CT scans are particularly useful for visualizing the condition of the lungs and can help in identifying characteristic signs of COVID-19, such as ground-glass opacities and bilateral pulmonary lesions, which are not always visible on standard X-rays \cite{kwekha2023coronavirus}.

Deep learning has emerged as a pivotal technology in medical image analysis, demonstrating remarkable success in enhancing the detection and diagnosis of various diseases, including COVID-19 \cite{li2020using}. By leveraging complex neural network architectures, deep learning models can automatically learn from vast amounts of medical imaging data, such as X-rays, CT scans, and MRI images, to identify intricate patterns and anomalies that may elude human experts. In the context of COVID-19, deep learning algorithms have been particularly instrumental in analyzing 3-D chest CT scans, enabling rapid, accurate identification of viral infections and assessment of disease severity \cite{fang2022annotation}. This capability has proved invaluable in managing the pandemic, as it assists healthcare professionals in making informed decisions quickly, optimizing patient management, and ultimately saving lives. 

The integration of deep learning in medical imaging for COVID-19 detection not only exemplifies the potential of AI in healthcare but also paves the way for its broader application in diagnosing a wide range of pathologies, promising a future where medical diagnostics are more efficient, precise, and accessible \cite{shi2020review}.
\begin{figure}[t]
  \centering
  \includegraphics[width=1\linewidth]{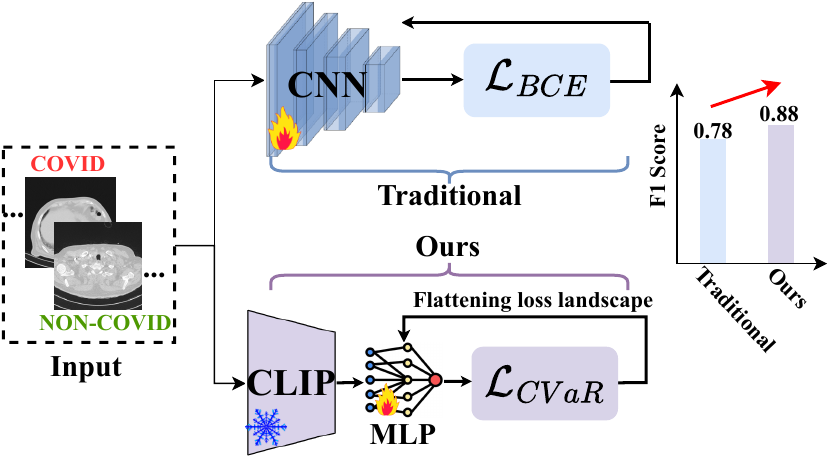}
  \vspace{-5mm}
  \caption{\textit{Comparison between our method with traditional method. \textbf{First row}: The traditional method trains a whole deep learning model (\eg, CNN) with a binary cross-entropy loss $\mathcal{L}_{BCE}$. \textbf{Second row}: Our method enhances COVID-19 detection by unitizing a frozen CLIP and a lightweight MLP classifier with Conditional Value at Risk
(CVaR) loss $\mathcal{L}_{CVaR}$ across a flattened loss landscape.}}
  \vspace{-5mm}
  \label{fig:introduction}
\end{figure}
However, the detection of COVID-19 using computational methods, especially through deep learning models, faces significant challenges that are critical to address for improving diagnostic accuracy and reliability. 
Firstly, the effectiveness of these models often hinges on the availability of extensive, high-quality datasets for training. However, real-world COVID-19 positive data is scarce, limiting the ability of these models to learn diverse manifestations of the disease \cite{wang2021methodp1}. This scarcity is compounded by the high variability in symptom presentation among different populations, making it difficult to develop models that are universally effective. Secondly, the issue of unlabeled data represents a substantial hurdle. Many datasets consist of images that have not been annotated with diagnostic outcomes, rendering them unusable for supervised learning methods without considerable effort to label them accurately.  Finally, improving detection performance with limited data while simultaneously enhancing model generalization requires innovative approaches.  Addressing these challenges is vital, as it directly impacts the models' ability to accurately diagnose COVID-19 across diverse global populations and varying stages of the disease, thereby playing a crucial role in managing the pandemic effectively \cite{wang2022methodp2}.

To address the challenges in COVID-19 detection due to limited and unlabeled data, existing methods have adopted several innovative strategies, yet they also encounter inherent limitations and gaps. To tackle the issue of limited data, techniques such as transfer learning have been widely used, where models pre-trained on vast, diverse datasets are fine-tuned using the smaller available COVID-19 datasets. This approach, however, may not always capture the unique characteristics of COVID-19-related anomalies due to the potential domain shift between the original and target datasets. For handling unlabeled data, semi-supervised and self-supervised learning methods have gained popularity. These methods leverage unlabeled data to learn general features or patterns, which can then be fine-tuned with a smaller labeled dataset for specific tasks. While effective to a degree, these methods can sometimes introduce biases or inaccuracies if the unlabeled data is not representative of the wider population or the specific nuances of COVID-19 pathology \cite{wang2022methodp3}. 

To improve generalization capabilities, existing approaches often employ data augmentation techniques to artificially expand the training dataset and introduce more variability, simulating a broader range of cases. Ensemble learning methods, which combine multiple models or predictions, are also used to enhance generalization. However, these strategies may still fall short when faced with highly diverse or novel cases outside of the training dataset's scope, highlighting a gap in the ability to robustly handle unseen variations. Overall, while existing methods have made significant strides in COVID-19 detection, challenges remain in ensuring high accuracy and generalizability across varied clinical settings and populations, underscoring the need for continuous innovation and validation in diverse real-world scenarios. These challenges also lead many competitions such as the 4$^{th}$ COV19D Competition \cite{kollias2024domain, kollias2023ai, arsenos2023data,kollias2023deep,kollias2022ai,arsenos2022large,kollias2021mia,kollias2020deep,kollias2020transparent}

\begin{figure*}[t]
    \centering
    \includegraphics[width=1\textwidth]{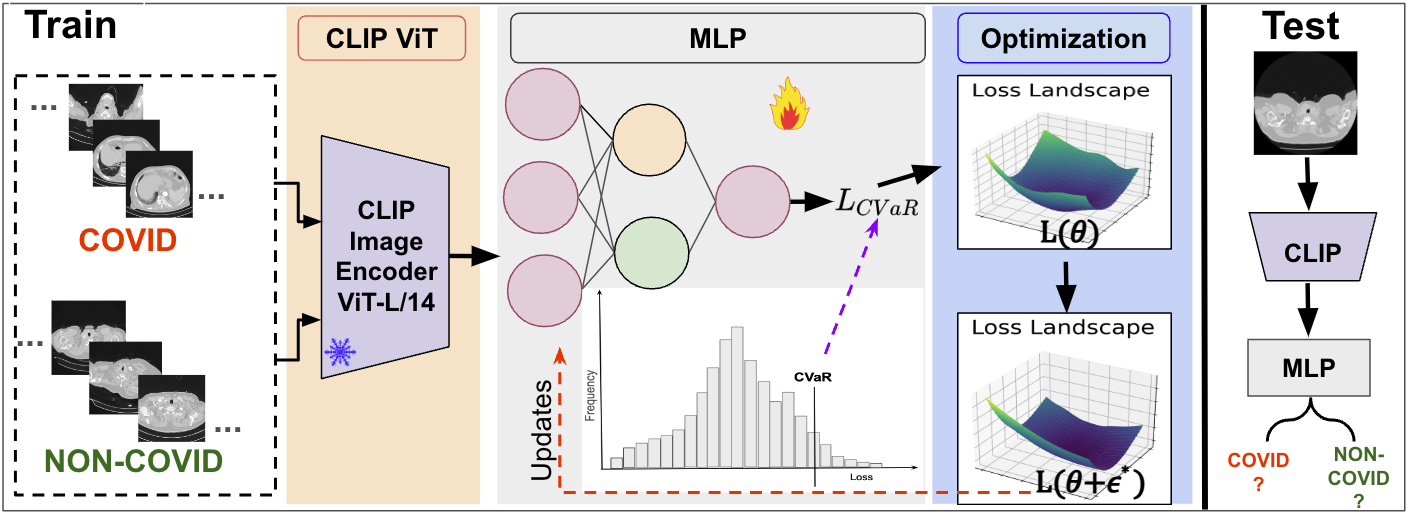}
    \vspace{-5mm}
    \caption{\textit{Overview of our proposed model using CLIP ViT for encoding the input images, an MLP module with robust CVaR loss, and an optimization step involving a flattened loss landscape for detecting COVID-19 cases apart from NON-COVID-19.}}
    \vspace{-4mm}
    \label{fig:SL}
\end{figure*}

In this work, we propose a novel framework as depicted in Fig.\ref{fig:SL}, comprising three modules: frozen Contrastive Language-Image Pre-training (CLIP)~\cite{radford2021learning} ViT as feature extractor, trainable classifier MLP, and optimization. This straightforward framework is used for both supervised and semi-supervised learning to detect COVID-19 from CT scan images. Specifically, 
for \underline{\textit{Supervised Learning}}, we leverage CLIP ViT-L/14~\cite{openclip2021} image encoder to capture high-level representations of the CT images. These representations are then fed into a 3-layer MLP trained to detect COVID-19 and non-COVID-19. We enhance the model's robustness and ability to focus on the most challenging cases by incorporating Conditional Value at Risk (CVaR) into the binary cross-entropy (BEC) loss. This is complemented by the optimization module, which enhances the model's generalizability by flattening the loss landscape. 
For \underline{\textit{Semi-Supervised Learning}}, we design a teacher-student framework in Fig.\ref{fig:SSL} to capitalize on the abundance of unlabeled data. The teacher model, after training on annotated data, assigns pseudo-labels to the unlabeled dataset. This process augments the training set, which is then used to train the student model. Through this transfer of knowledge, the student model is equipped to potentially surpass the teacher in detecting COVID-19.
Our contributions are summarized as follows:
\begin{enumerate}
    \item We propose the first lightweight detector for exposing COVID-19 based on labeled 3-D CT scans.
    \item We also propose a teacher-student framework for improving the COVID-19 detection performance by integrating unlabeled data.
    \item Our method outperforms state-of-the-art approaches, as demonstrated in extensive experiments on the COV19-CT-DB dataset.
\end{enumerate}

\section{Related Work}
\subsection{COVID-19 Detection}

According to ~\cite{shah2021comprehensive}, AI-empowered methods are employed for the detection of COVID-19 using medical images such as X-rays and CT scans. These include the preprocessing method and segmentation. Data preprocessing ~\cite{ahamed2021deep} involves resizing to 224x224 pixels, cropping for relevant regions, and sharpening filters to enhance edges, while data augmentation includes rescaling, zooming, flipping, and shearing to increase sample diversity and training robustness. The segmentation method proposed by ~\cite{lian2022lung} combines a DRD U-Net for image segmentation, integrating residual modules to enhance feature extraction, with a WGAN-based DNN classifier for efficient multi-class classification of COVID-19 images to train the classifier and optimize model parameters. Additionally, methods like transfer learning, fine-tuning, and novel architectures ~\cite{subramanian2022review} are employed in this domain.Transfer learning ~\cite{pratiwi2021effect} leverages pre-trained models like VGG-16, originally trained on large datasets like ImageNet, to classify COVID-19 CT-Scan images by fine-tuning some pre-trained layers, replacing the classifier layer, and utilizing features extracted by convolutional and pooling layers for classification.


\subsection{CLIP}


Contrastive Language-Image Pre-training (CLIP)~\cite{radford2021learning}, a simple yet effective pre-training paradigm, successfully introduces text supervision to vision models. It has shown promising results across various tasks in medical imaging (\ie, classification~\cite{lei2023clip, wang2022medclip}, detection~\cite{guo2023multiple}, and segmentation~\cite{liu2023clip}), attributable to its generalizability~\cite{radford2021learning}.

\section{Method}



\subsection{Supervised Learning}

\textbf{Feature Space Modeling}. We propose a simple procedure to tell apart COVID-19 CT scan series from non-COVID-19 based on features extracted from the image encoder of CLIP ViT L/14~\cite{openclip2021}. CLIP ViT is trained on an extraordinarily large dataset of 400M image-text pairs, so the high-level feature extracted from it is sufficient exposure to the visual world. Additionally, since ViT L/14 has a smaller starting patch size of $14 \times 14$ (compared to other ViT variants), we believe it can also aid modeling the low-level CT-scan slice details (\ie, Ground-Glass Opacities and Consolidation) needed for COVID-19 VS non-COVID-19 classification. Given a dataset $\mathcal{D}= \{(X_i, Y_i)\}_{i=1}^n$ with size $n$, where $X_i$ is the $i$-th CT scan slice and $Y_i \in \{0,1\}$ is the $i$-th sample label (0 means non-COVID-19, 1 means COVID-19). Feed CLIP visual encoder with dataset $\mathcal{D}$, and use its final layer to map the training data to their feature representations (of 768 dimensions). We get the resulting feature bank $\mathcal{C} = \{(F_i, Y_i)\}_{i=1}^n$ and further use the feature bank, which is our training set, to design an MLP classifier.


\textbf{MLP Classifier}. After constructing the feature bank, we use those feature embeddings to train a binary classifier to detect COVID-19 and non-COVID-19 CT scan slices. Our classifier is a straightforward 3-layer Multilayer Perceptron (MLP), and to foster a stable learning process and enhance the model's ability to generalize, we incorporate batch normalization after each linear transformation. This is followed by a ReLU activation, allowing for the model to capture intricate data patterns effectively. To further combat the risk of overfitting, a dropout layer is included following the activation function. 


\textbf{Objective Function}. To obtain a robust model, we apply a distributionally robust optimization (DRO) technique called \textit{Conditional Value-at-Risk} (CVaR)~\cite{rahimian2019distributionally, levy2020large, rockafellar2000optimization, hu2024outlier, ju2024improving, lin2024preserving, hu2023rank, hu2022distributionally, hu2021tkml, hu2022sum, hu2020learning}. By integrating CVaR into the binary cross-entropy (BEC) loss, the model is encouraged to pay more attention to the riskiest predictions. In the context of COVID-19 detection, these could be cases where the model is most uncertain and where misclassification could lead to the worst outcomes, such as failing to detect COVID-19 in patients with the disease. To this end, in what follows, we assume that $\mathcal{C} = \{(F_i, Y_i)\}_{i=1}^n$ consists of i.i.d. samples from a joint distribution $\mathbb{P}$, $F_i$ is the $i$-th data point's feature, and $Y_i$ is the $i$-th point's label. Given some variant of minibatch gradient descent, in the COVID-19 detection task, we are minimizing the empirical risk of the loss $\mathcal{L}_{avg}(\theta)= \frac{1}{n}\sum_{i=1}^{n}\ell(\theta, F_i, Y_i) \quad \text{for } \theta \in \Theta$,  instead of minimizing the true unknown risk $\mathcal{R}_{avg}(\theta) = \mathbb{E}_{(F,Y) \sim \mathbb{P}}[\ell(\theta; F, Y)]$, where $\ell$ is the individual loss function (\eg, binary cross-entropy) of the COVID-19 detection model, which has parameters $\theta$ for MLP.

However, the average loss is not robust to the imbalanced data, which is a common charismatic of the existing COVID-19 datasets. In addition, the training data distribution is usually not consistent with the testing data distribution, which is called the domain shift problem that widely exists in real-world COVID-19 detection scenarios. For example, the training set may be from one hospital but the test data may be from a different hospital. Therefore, we explore a DRO technique CVaR for handling imbalanced data, which can be formulated as: $\text{CVaR}_{\alpha}(\theta) = \inf_{\lambda \in \mathbb{R}} \left\{ \lambda + \frac{1}{\alpha} \mathbb{E}_{(F,Y)\sim P} \left[ \left(\ell(\theta; F, Y) - \lambda\right)_+ \right] \right\}$,
where $[a]_+ = \max\{0, a\}$ is the hinge function, the conditional value at risk at level $\alpha \in (0,1)$. As $\alpha \to 0$, we are concerned about minimizing the risk of `hard' samples (cases that are difficult to diagnose).  In contrast, as $\alpha \to 1$, it becomes minimizing the $\mathcal{R}_{avg}(\theta)$.
Inspired by~\cite{zhai2021doro}, we can minimize a loss function that aims to minimize an upper bound on the worst-case risk by employing the CVaR. In practice, we minimize an empirical version of $\text{CVaR}_{\alpha}(\theta)$. This gives us the following optimization problem:
\begin{equation}
\begin{aligned}
    \mathcal{L}_{CVaR}(\theta) =  \min_{\lambda \in \mathbb{R}}  \lambda + \frac{1}{\alpha n} \sum_{i=1}^{n} \left[ \ell(\theta; F_i, Y_i) - \lambda \right]_+.
\end{aligned}
\label{eq:learning objective}
\end{equation}

Suppose for a moment that we have obtained the optimal value of $\lambda^*$ in (\ref{eq:learning objective}), then the only training points that contribute to the loss are the `hard' ones with a loss value greater than $\lambda^*$, whereas the `easy' training points with low loss smaller than $\lambda^*$ are ignored. To this end, a robust model is obtained. These procedures are demonstrated in Fig.\ref{fig:SL}.

\begin{figure}[t]
    \centering
    \includegraphics[width=0.5\textwidth]{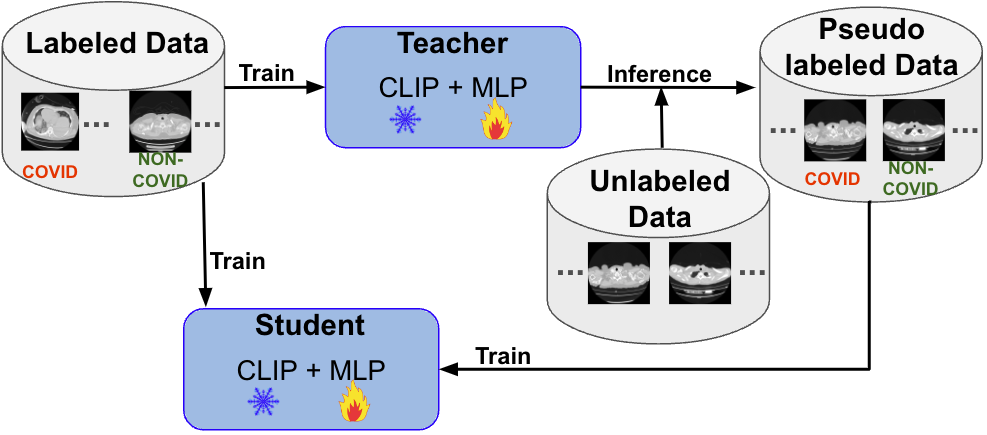}
    \vspace{-5mm}
    \caption{\textit{Diagrammatic representation of our robust model with teacher-student framework by leveraging unlabeled data for enhancing detection performance.}}
    \vspace{-6mm}
    \label{fig:SSL}
\end{figure}

\vspace{-1mm}
\subsection{Semi-Supervised Learning}
In addressing the challenge of limited labeled data for COVID-19 diagnosis from imaging, we propose a semi-supervised learning framework \cite{hu2022pseudoprop} that leverages knowledge distillation to utilize unlabeled data effectively. Our methodology comprises two primary phases: the teacher model training and the student model training. Note the model we use here is the same architecture as in supervised learning, both are fixed CLIP with a trainable 3-layer MLP. The whole training framework is illustrated in Fig.\ref{fig:SSL}. We explain each module as follows.

\textbf{Teacher Model Training}. The teacher model is first trained on a small set of labeled data. The aim of this phase is to develop a robust initial model that captures the high-level features and complexities associated with the diagnosis from the labeled dataset.

\textbf{Pseudo-Labeling with Teacher Model}. Once the teacher model is trained, it is employed to perform inference on a larger corpus of unlabeled data. The model's predictions are used to assign pseudo-labels to the unlabeled CT scan slices, creating a new set of training data that, while not verified by human experts, carries the inferred knowledge of the teacher model. These pseudo-labels are subject to uncertainty but offer a valuable starting point for expanding the training dataset beyond the limits of the labeled data.

\textbf{Student Model Training}. Subsequently, a student model is trained on a combination of the original labeled data and the newly created pseudo-labeled data. This process allows the student model to learn from both the ground truth in the labeled data and the nuanced patterns inferred by the teacher in the unlabeled data. The student model, through this extended training, is expected to outperform the teacher model by generalizing better to unseen data, thanks to the larger and more diverse training set it has been exposed to.

\subsection{Optimization}
Last, to further improve the detector's generalization capability, we optimize the COVID-19 detection model by utilizing the sharpness-aware minimization (SAM) method~\cite{foret2020sharpness} to flatten the loss landscape. Note that this optimization module can be used in both supervised learning and semi-supervised learning. As shown in Fig.\ref{fig:SL}, by utilizing such a technique, the model yields a more flattened loss landscape indicating a stronger generalization capability~\cite{lin2024preserving}. As a reminder, the model's parameters are denoted as $\theta$, flattening is attained by determining an optimal $\epsilon^*$ for perturbing $\theta$ to maximize the loss, formulated as:
\begin{equation}
    \begin{aligned}        
    \epsilon^*&=\arg\max_{\|\epsilon\|_2\leq \gamma}{\mathcal{L}_{CVaR}}\textbf{(}\theta+\epsilon \textbf{)}\\
    &\approx\arg\max_{\|\epsilon\|_2\leq \gamma}\epsilon^\top\nabla_\theta \mathcal{L}_{CVaR}=\gamma\sign(\nabla_\theta \mathcal{L}_{CVaR}),
    \end{aligned}
\label{eq:epsion_star}
\end{equation}
where $\gamma$ is a hyperparameter that controls the perturbation magnitude. The approximation is obtained using first-order Taylor expansion with the assumption that $\epsilon$ is small. The final equation is obtained by solving a dual norm problem, where $\sign$ represents a sign function and $\nabla_\theta \mathcal{L}_{CVaR}$ being the gradient of $\mathcal{L}_{CVaR}$ with respect to $\theta$. As a result, the model parameters are updated by solving the following problem:
\begin{equation}
    \begin{aligned}
        \min_{\theta} \mathcal{L}_{CVaR}\textbf{(}\theta+\epsilon^*\textbf{)}.
    \end{aligned}
\label{eq:sharpness}
\end{equation}
Perturbation along the gradient norm direction increases the loss value significantly and then makes the model more generalizable while detecting COVID-19. 

\begin{algorithm}[t]
    \caption{Optimization (CVaR+SAM)}\label{alg:Optimization}
    \SetAlgoLined
    \KwIn{A training dataset $\mathcal{C}$ with size $n$, $\alpha$, $\gamma$ max\_iterations, num\_batch, learning rate $\beta$}
    \KwOut{A robust COVID-19 detection model} 
    
    \textbf{Initialization:} $\theta_0$, $l=0$   

    
    \For{$e=1$ to \emph{max\_iterations}}{
    \For{$b=1$ to \emph{num\_batch}}{
    { \mbox{Sample a mini-batch $\mathcal{C}_b$ from $\mathcal{C}$}}
    
    Compute $\ell(\theta_l;F_i,X_i)$, $\forall (F_i,Y_i)\in \mathcal{C}_b$

    Use binary search to find $\lambda$ that minimizes (\ref{eq:learning objective}) on $\mathcal{C}_b$
    
    Compute $\epsilon^*$ based on Eq. (\ref{eq:epsion_star})

    Compute gradient approximation for (\ref{eq:sharpness})
    
    Update $\theta$: $\theta_{l+1}\leftarrow\theta_{l}-\beta \nabla_\theta \mathcal{L}_{CVaR}\big|_{\theta_l+\epsilon^*}$

    $l\leftarrow l+1$
    
    }
    }
    \Return{$\theta_{l}$}
\end{algorithm}

\textbf{End-to-end Training}. In practice, we first initialize the model parameters $\theta$ and then randomly select a mini-batch set $C_b$ from $C$, performing the following steps for each iteration on $C_b$ (see Algorithm~\ref{alg:Optimization}):
\begin{compactitem}
    \item Fix $\theta$ and use binary search to find the global optimum of $\lambda$ since (\ref{eq:learning objective}) is convex w.r.t. $\lambda$.
    \item Fix $\lambda$, compute $\epsilon^*$ based on Eq.~(\ref{eq:epsion_star}).
    \item Update $\theta$ based on the gradient approximation for (\ref{eq:sharpness}): $\theta\leftarrow\theta-\beta \nabla_\theta \mathcal{L}_{CVaR}\big|_{\theta+\epsilon^*}$, where $\beta$ is a learning rate.    
\end{compactitem}

\section{Experiments}
\subsection{Experimental Settings}
\subsubsection{Datasets}
The COV19-CT-DB dataset, as referenced in \cite{kollias2023deep}, forms the basis of our study, comprising 3-D chest CT scans. This collection encompasses 7,756 3-D CT scans, with 1,661 being COVID-19 positive and 6,095 being negative for COVID-19. It aggregates to approximately 2,500,000 images, of which 724,273 images are categorized under the COVID-19 class and 1,775,727 images under the non-COVID-19 class. Our analysis employs slices, meaning 2-D images derived from these scans, with each scan series containing 50 to 700 slices. Each slice has a resolution of 512$\times$512. This dataset also features in the 4$^{th}$ COV19D Competition as documented in \cite{kollias2024domain}. Based on the guidelines from \cite{kollias2024domain}, our approach involves: (a) using 703 3-D COVID-19 and 655 3-D non-COVID-19 CT scans from the dataset for supervised learning training, while reserving 170 3-D COVID-19 and 156 3-D non-COVID-19 CT scans for testing; (b) employing a semi-supervised learning framework, which includes CT scans obtained from various hospitals and medical facilities to ensure data diversity. In addition to the labeled scans used for supervised learning, our training set is augmented with 239 additional annotated 3-D CT scans (120 COVID-19 and 119 non-COVID-19) and 494 3-D CT scans without annotations, with the test set comprising 178 3-D CT scans (65 COVID-19 and 113 non-COVID-19). Further dataset details are elaborated in \cite{kollias2024domain}.

\subsubsection{Evaluation Metrics}
In line with the evaluation protocol established in \cite{kollias2024domain}, we adopt the macro F1 score as our primary metric for assessing the performance of all methods. This metric is essentially the unweighted mean of the F1 scores across different classes or labels, such as averaging the F1 scores for both the COVID-19 and non-COVID-19 categories. Given that our models are designed to classify individual images or slices rather than entire CT scans, we employ a majority voting strategy to aggregate the slice-level predictions into a singular diagnostic outcome for each CT scan. This approach allows us to compile the discrete predictions across all slices from a specific CT scan to arrive at a consolidated final diagnosis, thereby aligning our methodology with the comprehensive evaluation framework referenced in \cite{kollias2024domain}.

\subsubsection{Baseline Methods}
The baseline methods for our study are sourced from \cite{kollias2024domain}, which outlines two primary approaches:
(a) Within the supervised learning framework, we process input 3-D CT scans by applying padding to standardize their dimensions, which then proceed to a Convolutional Neural Network (CNN) segment for initial analysis. This CNN component is tailored to extract pivotal features, predominantly from lung areas, on an individual 2D slice basis. Following this feature extraction phase, a Recurrent Neural Network (RNN) sequentially processes the CNN-derived features, directing them through a Fully Connected (FC) layer and culminating in a softmax activation-based output layer dedicated to COVID-19 classification. A Dropout layer is integrated within this architecture to mitigate the risk of overfitting. This model is denoted as \textbf{CNN+RNN}.
(b) The semi-supervised learning strategy incorporates Monte Carlo Dropout to gauge uncertainty levels during the training phase with labeled data. This uncertainty measurement guides the annotation of unlabeled data, especially highlighting COVID-19 instances where the model demonstrates substantial confidence in its predictions. This technique is referred to as \textbf{Dropout}.
\begin{table}[t]
\centering
\begin{tabular}{c|c|c}
\hline
                                  & Method                       & `macro' F1 Score                       \\ \hline
                                  & CNN+RNN~\cite{kollias2024domain}                      & 0.780                                  \\
\multirow{-2}{*}{Supervised}      & \baseline{Ours} & \baseline{\textbf{0.886}} \\ \hline
                                  & Dropout~\cite{kollias2024domain}                      & 0.730                                  \\
\multirow{-2}{*}{Semi-Supervised} & \baseline{Ours} & \baseline{\textbf{0.734}} \\ \hline
\end{tabular}
\caption{\textit{Comparison with the baseline method in terms of `macro' F1 score under supervised and semi-supervised learning, respectively. The best results are shown in \textbf{Bold}.}}
\label{tab:results}
\vspace{-4mm}
\end{table}

\subsubsection{Implementation Details}
We adopt the CLIP framework, incorporating the Vision Transformer (ViT) as the image processing unit, specifically configured to the L/14 scale. This setup is augmented with three Multi-Layer Perceptron (MLP) layers, each consisting of 768 neurons, serving as our primary computational model. Our training protocol is executed with a batch size of 32, ensuring a precise management of samples during each training iteration.

The optimization process is facilitated by the Adam optimizer, which kicks off with an initial learning rate set at $\beta=1e-3$. Additionally, we employ a Cosine Annealing Learning Rate Scheduler to modulate the learning rate adaptively across the training duration, aiming to bolster the model's path to convergence. The tuning of hyperparameters involves adjusting $\alpha$ within the range of ${0.1, 0.2, 0.3, 0.4, 0.5, 0.6, 0.7, 0.8, 0.9}$ as delineated in Eq. (\ref{eq:learning objective}), alongside setting $\gamma$ to $0.05$ as specified in Eq. (\ref{eq:epsion_star}). These experiments are conducted using the PyTorch framework and leverage the computational prowess of an NVIDIA RTX A6000 GPU for training purposes.

\subsection{Results}
Table~\ref{tab:results} shows our results compared with the baseline method CNN+RNN~\cite{kollias2024domain} for two primary approaches: \textit{supervised learning} and \textit{semi-supervised learning}. It is clear that, in supervised learning, our method has superior COVID-19 detection ability compared to CNN+RNN~\cite{kollias2024domain}. It enhances the `macro' F1 score by \textbf{10.6\%}. When applied to semi-supervised learning, our method slightly outperforms the baseline. This is because our method simplifies the semi-supervised learning process by employing a direct pseudo-labeling approach without the added complexity of Monte Carlo Dropout for uncertainty estimation. Despite this simplification, our method achieves a higher `macro' F1 score. This indicates that the quality of pseudo-labels generated by our model is high, and our model is particularly effective at identifying and learning from the most informative unlabeled instances.

\subsection{Ablation Study}
\begin{figure}[t]
  \centering
  \includegraphics[width=1\linewidth]{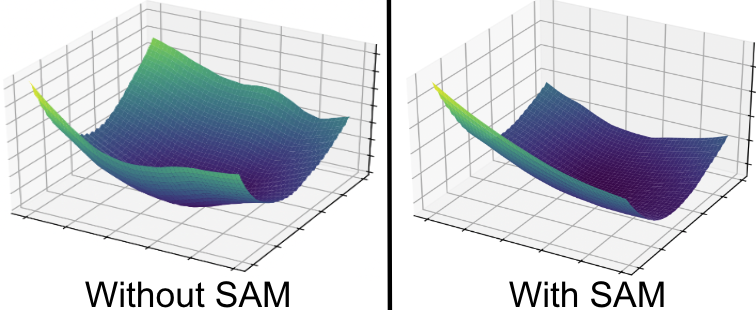}
  \vspace{-4mm}
  \caption{\textit{The loss landscape visualization of our proposed method without (left) and with (right) using the sharpness-aware minimization (SAM) method. The axis's scales are the same for both figures.}}
  \label{fig:sam}
\end{figure}

\begin{table}[t]
\centering
\begin{tabular}{c|c|c|c|c}
\hline
Method           & BCE & BCE+SAM & CVaR & \baseline{Ours(CVaR+SAM)} \\ \hline
`macro' F1 score & 0.868   & 0.874  &  0.877    & \baseline{\textbf{0.886}}          \\ \hline
\end{tabular}
\caption{\textit{An ablation study of our key components. `BCE' denotes cross-entropy loss, `BCE+SAM' represents BCE loss with the sharpness-aware minimization (SAM) optimization, `CVaR' and `Ours' represent Conditional Value at Risk without SAM and with SAM, respectively.}}
\label{tab:ablation}
\vspace{-3mm}
\end{table}

\textbf{Visualization of Loss Landscape}. Fig.\ref{fig:sam} visually illustrates the impact of incorporating SAM optimization in our proposed method. The loss landscape, without SAM, shows a more rugged and uneven loss surface. This unevenness can make the optimization process challenging, as it may lead to inconsistent generalization. In contrast, the right side reveals a much smoother loss landscape when SAM is applied. The smoother surface indicates a more robust model with parameters that generalize better to new data. The consistency in the loss surface with SAM also suggests that the optimization process is more straightforward, leading to improved learning during training. This visualization underscores the significance of the optimization module in our method for enhancing the detector's generalization. 

\textbf{Effects of CVaR and SAM}. The results in Table~\ref{tab:ablation} reveal the effects of CVaR technique and SAM optimization we applied in our proposed method. Compared with `BCE', `BCE+SAM' improves the `macro' F1 score by 0.6\%, `CVaR' enhances performance by 0.9\%, indicating the effectiveness of SAM optimization and CVaR loss, respectively. When we incorporate CVaR with SAM, it achieves the best performance surpassing `BCE' by 1.8\%. Overall, our method with both CVaR and SAM optimization yields the most substantial gains in `macro' F1 score.

\subsection{Sensitive Analysis}
\begin{figure}[t]
  \centering
  \includegraphics[width=0.8\linewidth]{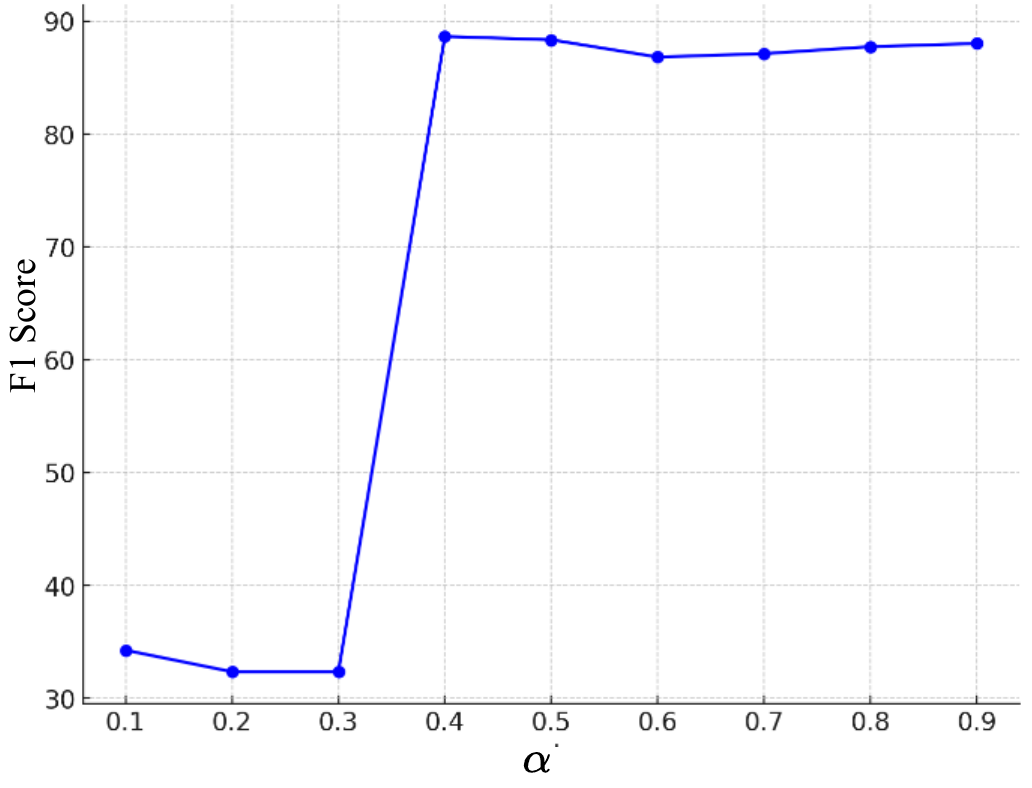}
  \vspace{-2mm}
  \caption{\textit{`Macro' F1 score to different $\alpha$ values.}}
  \vspace{-5mm}
  \label{fig:alpha}
\end{figure}

Fig.\ref{fig:alpha} shows the `macro' F1 score to different $\alpha$ values in Eq. (\ref{eq:learning objective}). The F1 scores are significantly lower when $\alpha$ is 0.1, 0.2, and 0.3. This is because the model focuses on the worst-case outcomes (extreme risks), it is either not predicting the positive class at all or predicting the negative class (F1 positive is 0 or F1 negative is 0). It shows a steep increase in the F1 Score at an $\alpha$ value of 0.4, indicating that the model's performance improves drastically when it moves away from the most extreme risk assessments to slightly more moderate ones. At this point, the balance between precision and recall that the F1 Score represents is much more favorable.

\section{Conclusion}
The current landscape of COVID-19 detection through deep learning models is faced with various challenges: requires large computational resources, the scarcity of high-quality, labeled datasets, and the abundance of unlabeled data. Addressing these issues, we introduce a streamlined detector that employs a frozen CLIP image encoder and a trainable MLP, augmented with CVaR and a loss landscape flattening strategy. The CVaR integration bolsters our model's robustness, while the loss flattening strategy enhances generalization. Moreover, our teacher-student framework adeptly leverages unlabeled data, ensuring effective model training even with limited labeled data. Experimental results showcase the superiority of our method compared with the baseline.

\textbf{Limitation.} A notable limitation of our study is that our proposed methodologies overlook the inherent correlations among CT images derived from the same 3-D CT scan. This oversight potentially results in the omission of valuable information that could enhance the diagnostic accuracy of our models.

\textbf{Future Work.} We plan to apply the CLIP text encoder by utilizing the medical diagnosis report data with the COVID-19 CT images to improve the detector's performance further.

\smallskip
\smallskip
\noindent\textbf{Acknowledgments.} This work is supported by the U.S. National Science Foundation (NSF) under grant IIS-2434967 and the National Artificial Intelligence Research Resource (NAIRR) Pilot and TACC Lonestar6.
The views, opinions and/or findings expressed are those of the author and should not be interpreted as representing the official views or policies of NSF and NAIRR Pilot.

\bibliographystyle{plain}
\bibliography{main}

\end{document}